\documentstyle[12pt,epsf,supercite]{article}
\begin{document}
\setcounter{totalnumber}{20}
\author{ Thomas V. Russo, Richard L. Martin, and\\P. Jeffrey
Hay\\Theoretical Division, MS B268\\Los Alamos National
	Laboratory\\Los Alamos, NM 87545\\LA-UR-94-4410}
\title{Application of Gradient-Corrected Density Functional Theory to
the Structures and Thermochemistries of
$\mbox{ScF}_3$, $\mbox{TiF}_4$, $\mbox{VF}_5$, and $\mbox{CrF}_6$}

%\titlepage
%\begin{singlespace}
\maketitle
%\end{singlespace}
\begin{abstract}

Density functional theory(DFT) and Hartree-Fock(HF) calculations are
reported for the family of transition metal fluorides $\mbox{ScF}_3$,
$\mbox{TiF}_4$, $\mbox{VF}_5$, and $\mbox{CrF}_6$.  Both HF and the
local-density-aproximation (LDA) yield excellent agreement with
experimental bond lengths, while the B-LYP gradient-corrected density
functional gives bond lengths $0.04-0.05$ \AA\ too long.  An
investigation of various combinations of exchange and correlation
functionals shows that, for this series, the origin of this behavior
lies in the Becke exchange functional.  Much improved bond distances
are found using the hybrid HF/DFT functional advocated by Becke. This
approximation also leads to much improved thermochemistries.  The LDA
overestimates average bond energies in this series by $30-40$
kcal/mol, whereas the B-LYP functional overbinds by only $\sim8-12$
kcal/mol, and the hybrid HF/DFT method overbinds by only $\sim 2$
kcal/mol.  The hybrid method predicts the octahedral isomer of
$\mbox{CrF}_6$ to be more stable than the trigonal prismatic form by
$14$ kcal/mol.  Comparison of theoretical vibrational frequencies with
experiment supports the assignment of an octahedral geometry.

\end{abstract}
\pagebreak

% smear a big gray DRAFT across the pages.
%\special{!userdict begin /bop-hook{gsave 200 30 translate 65 rotate
%/Times-Roman findfont 216 scalefont setfont 0 0 moveto 0.9 setgray
%(DRAFT) show grestore}def end}

\section{Introduction}

The new generation of ``gradient corrected'' density
functionals have been shown to yield remarkably accurate thermochemistries
for organic systems\cite{johnson,AW}.
Applications to inorganic complexes are less extensive,
but those studies which have been reported are very
encouraging\cite{Ziegler,DWL,Deeth,Li,bausch}.
We have recently developed a DFT code
\cite{RMH} and applied it to the excitation and ionization energies of the
atoms of the first transition series. The Becke exchange functional
\cite{becke88}
coupled with the Lee-Yang-Parr correlation functional \cite{LYP}, the
(B-LYP) functional, gave exceptional results for the ionization
energies (mean errors of $\sim0.1$ eV) and acceptable results for excitation
energies (mean errors of $\sim0.5$eV). We were therefore interested
in the predictions of gradient-corrected DFT for a ``simple'' series of
ionic complexes,  the transition metal fluorides:
$\mbox{ScF}_3$, $\mbox{TiF}_4$, $\mbox{VF}_5$, and $\mbox{CrF}_6$.
Experimental geometries and bond energies are in hand for the
first three members of this series, thereby allowing a study of
the accuracy of the results as a function of basis set and functional,
while the last one,
$\mbox{CrF}_6$, has been the subject of considerable controversy in
recent years. On the experimental side, the debate centers about
whether the vibrational spectrum observed by
Hope, et al.\cite{HJLO,HLO},
and used by them to assign $O_h$ symmetry to
$\mbox{CrF}_6$,  indeed corresponds to $\mbox{CrF}_6$, or is actually the
signature
of $\mbox{CrF}_5$\cite{JW,JMWJB}.
The debate among theorists centers
around the relative stability of the $O_h$
isomer relative to a trigonal prismatic geometry of $D_{3h}$ symmetry
\cite{KAE,MW,PR,NFHG,MMQ}. Early calculations preferred the $D_{3h}$
structure but the most recent and extensive calculations suggest that
the $O_h$
isomer is the more stable.

We present here the results of calculations on this series using both
a local density approximation(LDA) and the gradient-corrected B-LYP
approximation.  The local density approximation comprises the Slater
exchange functional and the Vosko-Wilks-Nusair parameterization of the
electron gas correlation energy (S-VWN).  Hartree-Fock calculations
were also performed for comparision.  Our results show that the B-LYP
approximation yields reasonable binding energies for the transition
metal fluorides using modest basis sets, but tends to overbind by
8--10 kcal/mol per bond.  We have also investigated a hybrid
HF/DFT method due to Becke
(the Becke3LYP method) \cite{becke3}, and have found this to yield
very good results
(errors of the order of $\sim 2$ kcal/mol per bond).
The calculations support the claims
that the octahedral structure of $\mbox{CrF}_6$ is of lower energy
than the trigonal prismatic structure, our best calculation
(Becke3LYP) favoring the octahedral structure by 14 kcal/mol.

\section{Method}

The DFT calculations were carried out using the MESA suite of
programs\cite{MESA93}, into which we recently added DFT
modules\cite{RMH}.  The Kohn-Sham procedure used is strictly analogous
to traditional Hartree-Fock calculations, with the replacement of the
analytical exchange matrix with an exchange-correlation matrix
calculated by numerical integration \cite{johnson}.  Analytic
gradients of the energy were computed including the derivatives of the
quadrature weights
\cite{johnson}.

The ``Standard Grid-1'' of Gill et al.\cite{sg1} is used on the
fluorines; on the metals we use a simple grid of 100 concentric
spheres, with radii chosen according to an Euler-MacLaurin
scheme\cite{murray}, on which are positioned quadrature points chosen
according to the scheme of Lebedev\cite{lebedev}, using the 194 point
(23rd order) formul\ae.  This 100-point radial grid for the transition
metal has been shown to provide atomic energies accurate to
$10{^{-4}}$ a.u. in a previous study \cite{RMH}.  It is much larger
than the fluorine grids and could be made significantly more
economical with little loss in accuracy by applying a ``grid pruning''
technique\cite{sg1,PBS,baker}.

Most atomic calculations were performed using a spin-restricted
open-shell procedure as described in a previous work\cite{RMH}.

The hybrid HF/DFT, Becke3LYP, calculations were performed with
the Gaussian 92/DFT programs\cite{gaussian92}. A 100 point radial
grid and 194 point angular grid was used on each atom. Bond energies
were obtained using atomic energies calculated with a
spin-unrestricted Kohn-Sham procedure.

\subsection{Basis sets}

Our initial studies on these molecules used a general contraction of the
$(14s9p5d)$ Cartesian-Gaussian basis set of Wachters\cite{wachters}
augmented with the diffuse $d$ function of Hay\cite{hay-dif-d}. As
described previously\cite{RMH}, we have used Wachters' Hartree-Fock
coefficients in our contraction, and contract the inner parts of the
$1s$, $2s$, $3s$, $2p$, $3p$, and $3d$ orbitals to obtain a $[6s5p3d]$
set using the general contraction scheme of
Raffenetti\cite{gen-contract}. The more diffuse primitives of
each space are left free.  This contracted set gave B-LYP total energies
for Sc-Cr to within $10{^{-3}}$ a.u. of the completely uncontracted results,
and
excitation and
ionization energies to within $0.01$eV\cite{RMH}.
With this metal basis set, we then examined various
levels for the fluorine basis set in $\mbox{CrF}_6$. Table~\ref{tb:fbas}
reports the
optimal geometrical parameters, average bond energies,
and relative stability for the ${O_h}$
and ${D_{3h}}$ conformers of $\mbox{CrF}_6$ in the B-LYP approximation.
Examination of Table~\ref{tb:fbas} reveals that inclusion of the $d$
polarization function (6-31G*) decreases the optimum bond length in both
the ${O_h}$ and ${D_{3h}}$ conformers by some 0.02 \AA. Consistent with this
shorter bond length, the average bond energy increases by $\sim5$ kcal/mol
in both conformers.  The relative energy is only slightly changed, the ${O_h}$
conformer being more stable by $\sim9$ kcal/mol. Because these species
are expected to be ionic in character, calculations were performed in which
diffuse $s$ and $p$ functions were added to the fluorine atom (6-31G+).
Relative to the double-zeta 6-31G basis, this modification increased
the bond lengths by $0.01$ \AA, and decreased
the average bond energy by $\sim5$ kcal/mol. The addition of
the diffuse functions had a more pronounced influence on the energy difference
than did inclusion of the polarization function; the ${O_h}$ form is $16.6$
kcal/mol more stable than the $D_{3h}$ form in the (6-31G+) basis.
 A final calculation in which
both diffuse and polarization functions were added to the fluorines (6-31G+*)
yields a nearly identical relative stability of $16.3$ kcal/mol.  In the
calculations to follow, then, the (6-31G+*) basis is used for the ligands.

There were additional issues regarding the transition metal basis set which we
wished to address before examining the series.
First of all, folklore has it that Wachters' primitive
set optimized for the atoms is sometimes inadequate in the valence 4s/p
region in a molecule.  Specifically, the primitive $s$ space may be too
diffuse, and the functions provided by Wachters' to describe the $4p$
atomic orbital are also more diffuse than those which might
be needed to act as a polarization function in a molecule.
Preliminary tests were therefore run on $\mbox{ScF}_3$ and $\mbox{TiF}_4$
in which the two most diffuse of Wachters' $s$ primitives were replaced by
tighter functions chosen to make the set more even-tempered.  For Sc the new
exponents were .12 and .04, for Ti they were .2 and .07.
For $\mbox{ScF}_3$ the change in total energy with the modification of the
$s$ set was only $\sim0.25$ kcal/mol.
Addition of a diffuse $p$ function, ${\alpha = 0.15}$,
changed the total energy of $\mbox{ScF}_3$ by $\sim1$ kcal/mol.
Similar results were found for $\mbox{TiF}_4$.
Modification of the metal $s$ and $p$ set thus appears to be a minor
effect in this series,
presumably because of the ionic character of
the bonding.

A second issue concerns the relative importance of $f$ functions in
DFT calculations.  Previous research using CI and coupled-cluster
techniques discovered a profound effect on the relative stability when
$f$ polarization functions were included on the metal.  We found that
addition of a metal $f$ function, optimized in the B-LYP calculation,
had a relatively minor effect on the energy. In both $\mbox{ScF}_3$
and $\mbox{TiF}_4$, addition of an optimized $f$ function changed the
total energy by $\sim5$ kcal/mol, or $\sim1$ kcal/mol in the average
bond energy.  In the discussion to follow, we compare results using
the $[6s5p3d]$ contraction of Wachters' original basis and this basis
augmented with an $f$ function optimized in the B-LYP calculation
$[6s5p3d1f]$.  These optimized $f$ exponents were found to be 0.4 for
Sc, and 0.5 for the other three molecules.

\section{Results and Discussion}

\subsection{$\mbox{ScF}_3$, $\mbox{TiF}_4$, and $\mbox{VF}_5$}

Table~\ref{tb:results1} compares theoretical and experimental bond
lengths for $\mbox{ScF}_3$, $\mbox{TiF}_4$ and $\mbox{VF}_5$\cite{DH}.
The two major headings refer to the results using the $[6s5p3d]$ and
$[6s5p3d1f]$ metal atom bases.  It can be seen that the  additional
metal $f$ function has little effect on the equilibrium distances, regardless
of the theoretical method used. If any trend can be observed, it is a tendency
for the $f$ function to decrease the bond lengths slightly ($\sim0.01$\AA).
Concentrating on the $[6s5p3d1f]$ results, HF theory and the S-VWN aproximation
are generally in good agreement with experiment. An exception occurs for
$\mbox{ScF}_3$ where HF theory underestimates the bond length by $0.05$\AA,
with S-VWN being $0.08$\AA\ too short. This exception was also noted in the
S-VWN calculations of Sosa et al.\cite{Sosa}
  Although the basis sets, grids, and other
details of our calculations differ from theirs, the agreement between the
present S-VWN results and this earlier research is excellent.

Perhaps most disappointing in Table~\ref{tb:results1} is the tendency
of the B-LYP approximation to overestimate the bond lengths.  For
$\mbox{TiF}_4$ and $\mbox{VF}_5$, where HF and S-VWN are in good
agreement with experiment, the B-LYP results are $\sim0.04-0.05$\AA\
longer than experiment.  Similar behavior was obtained with B-LYP
in a series of molecules comprised of first row atoms
\cite{johnson},
although the error was smaller, of the order of 0.02\AA.
In the transition-metal fluoride series studied here, the origin
of this behavior is associated with the Becke exchange functional.
This can be seen from Table~\ref{tb:results3}
where the equilibrium distances predicted by various functional
combinations are presented.  Note that both the B-VWN and B-LYP
approximations give distances which are too long, while S-VWN and
S-LYP give distances in good agreement with experiment.
The reason for this behavior is unknown; as is shown
below, the hybrid HF/DFT version suggested by Becke largely remedies
the problem\cite{becke3}.

Table~\ref{tb:results2} compares average bond energies,
$\overline{E}$, for the various methods with experiment.  The
experimental values in Table~\ref{tb:results2} were obtained from
enthalpies of formation at $298.15 {}^\circ\mbox{K}$\cite{barin}.  The
theoretical quantities reported are obtained by subtracting the energy
of the molecule from the sum of the energies of the atoms, and
dividing by the number of fluorines.  In order to compare the
theoretical results directly with experiment, they should be corrected
for the zero-point contribution and for the difference in enthalpies
of formation between 298 and 0${}^\circ$K.  In the case of
$\mbox{TiF}_4$, this thermodynamic information is available from
experiment\cite{janaf}.  To the experimental heat of atomization at
298${}^\circ$K, the zero-point contribution and differences in
enthalpy for 0${}^\circ$K, +5.5 kcal/mol and -3.5 kcal/mole,
respectively, should be added to obtain a ``corrected'' experimental
value.  This ``corrected'' value thus increases by 2.0 kcal/mol, or
0.5 kcal/mol/bond for $\mbox{TiF}_4$.  As a result the average bond
energy would be 140.4 instead of 139.9 kcal/mol, which can be compared
directly with the theoretical value of 149.5 kcal/mol for BLYP.  While
vibrational frequencies are known for $\mbox{VF}_5$, the other
information needed to make corrections for the other molecules is not
available.  We estimate that these corrections would range from 0.3 to
0.7 kcal/mol for the average bond energies as one goes from
$\mbox{ScF}_3$ to $\mbox{CrF}_6$.  These corrected estimates appear
parenthetically in Table~\ref{tb:results2}

\begin{figure}[h]
\leavevmode
\epsffile{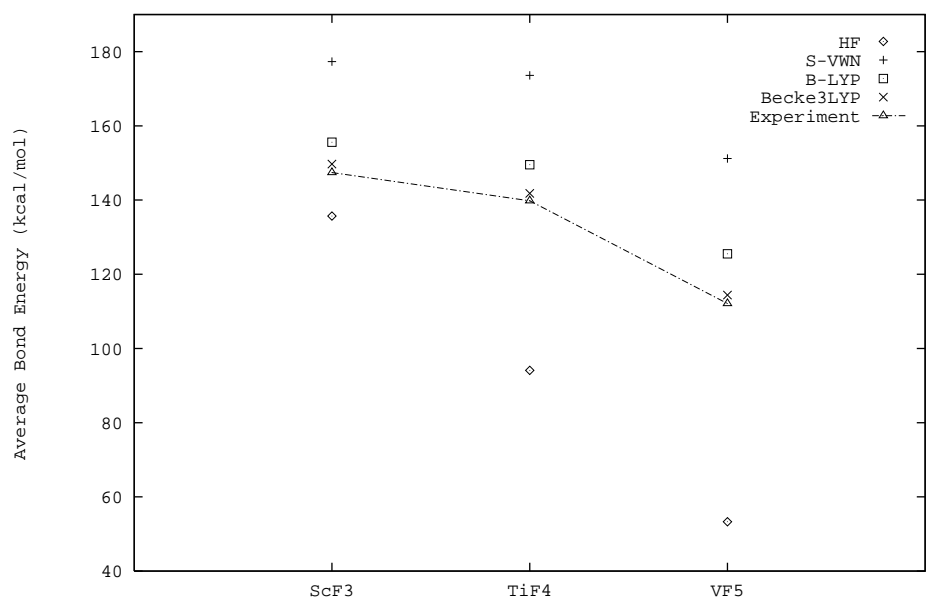}
\caption{
The average bond energy in kcal/mol for the transition metal fluoride
series for HF($\Diamond$), S-VWN(+), and B-LYP($\Box$), Becke3LYP
($\times$) and experiment($\triangle$).}
\label{gf:bonderr}
\end{figure}

The first point to notice is that the $f$-functions have a minor influence
on the average bond energy. They contribute $\sim2$ kcal/mol per bond in the
DFT calculations, and somewhat more for the HF calculation,
up to $\sim4$ kcal/mol in $\mbox{VF}_5$.  As expected, the HF bond energies
are in poor agreement with experiment.  The error per bond grows quickly
as the number of bonds increases;
the error is $\sim60$ kcal/mol per bond in
$\mbox{VF}_5$.  This is shown graphically in Figure~\ref{gf:bonderr}.  On the
other hand,
the error in the S-VWN DFT approximation is more constant as the
size of the molecule increases.  The S-VWN approximation overbinds the
molecule, by $30$, $33$, and $39$ kcal/mol per bond in $\mbox{ScF}_3$,
$\mbox{TiF}_4$,
and $\mbox{VF}_5$, respectively.  The B-LYP bond energies are in much better
agreement with experiment.  Like S-VWN, this approximation also overestimates
the bond energy, but by only $8$, $9$, and $13$ kcal/mol, respectively.
This error is somewhat larger than the typical error observed in studies of
organics, but it should be kept in mind that quite extensive theoretical
treatments including very large one- and many-electron bases would be necessary
in order to obtain this degree of accuracy in
conventional HF-based correlation schemes.

It is sometimes argued that more accurate bond energies are obtained
when using spin-unrestricted Kohn-Sham (UKS) atomic results, as
opposed to the spin-restricted open-shell method used here. Since the
DFT approximations generally overbind the molecule, and the UKS method
will always yield atomic energies identical to or lower than the ROKS
method, the use of UKS atomic energies will certainly decrease the
bond energies, thereby improving the apparent agreement with
experiment. In order to investigate the magnitude of this effect, we
calculated UKS atomic energies using Gaussian92/DFT\cite{gaussian92}.
In the $[6s5p3d]/[6-31G+*]$ basis set, the average bond energies were
153.4, 146.8 and 122.3 for $\mbox{ScF}_3$, $\mbox{TiF}_4$ and
$\mbox{VF}_5$.  The difference between our ROKS and these UKS results
is small ($\sim 1$ kcal/mole per bond).

Becke\cite{becke3} has argued that the residual tendency of the
gradient-corrected DFT methods to overbind stems from the exchange
functional and has introduced a hybrid method in which a component
of the ``exact'' Hartree-Fock exchange is retained in the energy
expression.  We have therefore also examined a variant of this hybrid
approach, the Becke3LYP method.

The $[6s5p3d1f]/[6-31G+*]$ basis set was used, giving the results
shown in Table~\ref{tb:b3lyp}.  The results are in very good agreement
with experiment: the difference in the average bond energies between
the raw Becke3LYP results and the ``corrected'' experimental values is
2.0, 1.4 and 1.7 kcal/mol per bond.  Bond lengths are also
significantly improved over the B-LYP approach.  They are
approximately 0.01\AA\ too long, except in the case of $\mbox{ScF}_3$,
where the bond length is underestimated by 0.05\AA.  Given the success
of the method for the other molecules, this experimental determination
should probably be revisited.

\subsection{$\mbox{CrF}_6$}

Before discussing the present results for $\mbox{CrF}_6$, a brief review
of the history of the problem is useful.
In 1963, Hellberg, M\"uller and Glemser\cite{HMG} reported the
preparation of a volatile yellow substance which they identified as
$\mbox{CrF}_6$.  On the basis of vibrational data,
Hope, et al.\cite{HJLO} assigned
octahedral $({O_h})$ symmetry to the molecule.
The first {\em ab initio} theoretical treatment of $d^0$ $ML_6$ complexes
was reported by Kang, Albright and Eisenstein in 1989 \cite{KAE};
these studies predicted octahedral geometry for $\mbox{CrF}_6$ at the
MP2/minimum basis set level of theory.
This prediction was countered by the calculations of
Marsden and Wolynec\cite{MW} who concluded
that $\mbox{CrF}_6$ was probably trigonal prismatic $({D_{3h}})$
based on the predictions of coupled-cluster theory with double substitutions
(CCD) in a double-zeta basis set.
Jacob and Willner \cite{JW}
carried out new experiments and argued that
$\mbox{CrF}_6$ had not yet been prepared, and that
earlier experiments had merely produced $\mbox{CrF}_5$.  Hope, Levason
and Ogden \cite{HLO} responded with  a more detailed analysis of their data
and again concluded that $\mbox{CrF}_6$ had octahedral geometry.

More extensive theoretical treatments then came into play.
Pierloot and Roos \cite{PR} performed CASPT2 (complete active space
with second-order M\o ller-Plesset perturbation theory) calculations
in a large basis set, and found that the octahedral isomer was 50
kcal/mol lower in energy than the prismatic isomer.  Neuhaus, Frenking,
Huber and Gauss
\cite{NFHG}
discovered two interesting facts. The first is a large differential effect
in the SCF energies when $f$ functions are included in the
Cr basis set. SCF calculations without $f$ functions showed the
prismatic isomer to be preferred, but with $f$ functions the
octahedral isomer was favored by 11 kcal/mol. Neuhas, et al.\ then explored a
hierarchy of coupled cluster expansions with a smaller double-zeta basis and
discovered a second interesting point:
a large differential effect in the correlation energies
of the two isomers associated with the
contribution of triple excitations.  In their double-zeta basis it
was nearly as large as the differential effect of $f$ functions at the SCF
level.
By combining these two results, Neuhas et al.\ concluded that the octahedral
isomer was lower in energy by some 20 kcal/mol.
Meanwhile, back in the lab, a
1992 experimental paper from Jacobs, M\"uller, Willner, Jacob and
B\"urger \cite{JMWJB} once again argued that $\mbox{CrF}_6$ had not in fact
been
prepared.  Most recently, in 1994, Marsden, Moncrieff and Quelch \cite{MMQ}
reported correlated calculations using CISD, CCSD and CCSD(T) methods
with a large basis including $f$ functions on the metal. Their work confirmed
the sensitivity of the results to metal $f$ functions and triple excitations.
Their most extensive calculation favored the octahedral isomer by $14.2$
kcal/mol.
Thus, at the present time, theory is converging to the position
that the octahedral form is more stable than the prismatic. The magnitude
of the difference is still not settled, particularly given the large effect
of triple excitations on the relative stability.

The results of the B-LYP calculations for $\mbox{CrF}_6$ are presented in
Table~\ref{tb:results1} and Table~\ref{tb:results2}, and those using the
Becke3LYP approximation in Table~\ref{tb:b3lyp}.  Once again,
the Becke3LYP method yields shorter bond lengths and smaller binding
energies than B-LYP. The Becke3LYP equilibrium
bond distances for both the ${O_h}$ and ${D_{3h}}$ forms are similar,
1.73\AA\ and 1.74\AA\, respectively.
The optimum bond angle in the ${D_{3h}}$ isomer,
i.e. that between a three-fold axis and a Cr-F bond,
was found to be $50.45^\circ$,  in good agreement with earlier HF calculations
which found $50.5^\circ$.  The average bond energy follows the trend
established in the earlier members of the series of decreasing strength
with increasing number of bonds.
The difference in energy
between the two isomers
is $13.8$ kcal/mol.  This result is very close to the latest
CCSD(T) calculations of Marsden, et al.\ who found 14.2 kcal/mol.

Since the experimental efforts at identifying the molecule have
focussed on the vibrational spectrum, we present in
Table~\ref{tb:freqs} the B-LYP harmonic frequencies of the $O_h$
isomer.  The two infrared active modes occur at $329$ and
$717\mbox{cm}^{-1}$.  The former is in excellent agreement with the
experimental bend observed\cite{HLO} at $332\mbox{cm}^{-1}$.  The
latter is in reasonable agreement with the intense mode observed at
$760 \mbox{cm}^{-1}$.  Although there are not many cases reported yet,
it appears that the gradient-corrected approximations tend to
underestimate metal-ligand stretching frequencies. This would be
consistent with the observed tendency of B-LYP to overestimate bond
lengths.  As discussed above, the Becke3LYP approximation
significantly improved both the bond lengths and the bond energies
relative to B-LYP; we therefore also examined the vibrational
frequencies predicted by Becke3LYP (Table~\ref{tb:freqs}).  The
theoretical predictions of 768 and 336 $\mbox{cm}^{-1}$ are in
excellent agreement with the experimentally observed bands.
Furthermore, the ratio of the IR intensities of the Becke3LYP modes is
0.081, which agrees with the corresponding ratio in the experiments of
Jacob and Willner\cite{JW}, 0.080.  These calculations strongly
support both the conclusion that $\mbox{CrF}_6$ has been prepared and
the assignment of $O_h$ symmetry.

\section{Conclusions}

As has been noted many times before\cite{andzelm}, an advantage of the DFT
approach
is that the density, and hence the
energy, converges much more rapidly with basis
set than does the correlation energy in configuration-interaction or
coupled-cluster approaches. A case in point in the present work
is the relative insensitivity of the results to the presence of $f$ functions
on the metal.  This can be contrasted with the research reviewed
above.  In Hartree-Fock based correlation approaches the $f$
functions must provide for angular correlation among the $d$ electrons,
while in DFT calculations they play the role of polarization functions only.
The small effect they have on the structure and bond energies in this series
is presumably a reflection of the ionic character of the bonding. In the ionic
limit, the metal density is roughly spherical, and the dominant polarization
effects should come from $p$ functions.

The bond energies provided by the B-LYP approximation are in fairly good
agreement with experiment.  The decrease in average bond energy down the
series is faithfully reproduced, and the absolute values are within $\sim10$
kcal/mol of experiment even in this relatively modest double-zeta-polarization
basis set. The B-LYP distances, however, appear to be consistently too long
by $\sim0.04$ \AA, with the exception of $\mbox{ScF}_3$.  The anomalous
behavior of $\mbox{ScF}_3$ suggests a problem with the
experimental data.

Both the bond distances and energies are greatly
improved in the hybrid HF/DFT Becke3LYP approach.
Distances are within 0.01\AA\ and average bond energies within 2 kcal/mole
per bond of experiment.  This success is impressive, especially considering
that the molecules studied here are quite different from those used by
Becke to generate his fit.  We note that Ricca and Bauschlicher\cite{bausch}
have recently found that the Becke3LYP approximation provides excellent
thermochemistries in the series $\mbox{Fe(CO)}_n^+$.  The bonding in these
carbonyls is certainly quite different from the ionic limit studied in
the current contribution; the approach appears very promising given the
small sample of transition metal complexes studied thus far.

Finally, our B-LYP calculations suggest that the octahedral form of
$\mbox{CrF}_6$ is preferred over the $D_{3h}$ form by $\sim17$
kcal/mol; this number is reduced to 14 kcal/mol in the Becke3LYP
approximation.  The Becke3LYP IR-active vibrational frequencies and
relative intensities computed for the $O_h$ structure are in excellent
agreement with the experimental infrared spectrum and support the
original assignment of Hope, et al.

\section{Acknowledgments}
This work was sponsored by the U.S. Department of Energy through the
LDRD program at Los Alamos.

\pagebreak

\begin{table}[ht]
\begin{center}
\begin{tabular}{lrrrr}\hline\hline
 Basis&6-31G&6-31G*&6-31G+&6-31G+*\\\hline
$r({O_h})$&1.78&1.76&1.79&1.75\\
$r({D_{3h}})$&1.79&1.77&1.80&1.78\\
$\overline{E}({O_h})$&93.1&98.3&88.2&92.6\\
$\overline{E}({D_{3h}})$&91.5&96.8&85.4&89.9\\
$E({D_{3h}}$-${O_h})$&9.5&9.0&16.6&16.3\\\hline\hline
\end{tabular}
\end{center}
\caption{Equilibrium bond lengths, average bond energies, and conformer
stability  for $\mbox{CrF}_6$. The columns denote various F bases
used in conjunction with the $[6s5p3d]$ contraction of the
Wachters-Hay primitive set for Cr ($r$ in \AA, $E$ in kcal/mol).}
\label{tb:fbas}
\end{table}

\begin{table}[ht]
\begin{center}
\begin{tabular}{llr|r|r||r|r|r||r}  \hline\hline
&&\multicolumn{3}{c}{$[6s5p3d]$}&
  \multicolumn{3}{c}{$[6s5p3d1f]$}&
  \multicolumn{1}{c}{Expt}\\
\multicolumn{2}{l}{Molecule}
                        &HF    &S-VWN &B-LYP &HF    &S-VWN& B-LYP&    \\
\hline
$\mbox{ScF}_3$ &        &1.860 &1.831 & 1.868&1.858 &1.829& 1.865&1.91\\
$\mbox{TiF}_4$ &        &1.747 &1.745 & 1.781&1.743 &1.740& 1.776&1.745\\
$\mbox{VF}_5$  &$r_{ax}$&1.728 &1.747 & 1.786&1.718 &1.738& 1.778&1.734\\
               &$r_{eq}$&1.685 &1.709 & 1.748&1.681 &1.704& 1.744&1.703\\
$\mbox{CrF}_6$ &${O_h}$   &    &      & 1.753&     &      & 1.765&    \\
               &${D_{3h}}$&    &      & 1.782&     &      & 1.775&    \\
\hline\hline
\end{tabular}
\end{center}
\caption{Optimum bond lengths(\AA) from S-VWN, B-LYP and HF
calculations on the transition metal fluoride series. The calculations
were carried out with the 6-31G+* basis on the fluorine atoms and either
the $[6s5p3d]$ or the $[6s5p3d1f]$ contraction described in the text.
The experimental results are summarized in Ref.~\protect\citenum{DH}.
For $\mbox{CrF}_6(D_{3h})$, the optimum bond angle between a
three-fold rotation axis and the Cr-F bond is $50.44^\circ$ and
$50.45^\circ$ in the $[6s5p3d]$ and $[6s5p3d1f]$ metal bases,
respectively.}
\label{tb:results1}
\end{table}

\begin{table}[ht]
\begin{center}
\begin{tabular}{ll|r|r|r|r||r}  \hline\hline
&&\multicolumn{4}{c}{$[6s5p3d]$}&
  \multicolumn{1}{c}{Expt}\\
\multicolumn{2}{l}{Molecule}
                        &S-VWN &S-LYP &B-VWN    &B-LYP   &    \\  \hline
$\mbox{ScF}_3$ &        &1.831 &1.826 &1.874    &1.868   &1.91\\
$\mbox{TiF}_4$ &        &1.745 &1.739 &1.787    &1.781   &1.745\\
$\mbox{VF}_5$  &$r_{ax}$&1.728 &1.740 &1.794    &1.786   &1.734\\
               &$r_{eq}$&1.685 &1.703 &1.756    &1.748   &1.703\\ \hline\hline
\end{tabular}
\end{center}
\caption{Optimum bond lengths(\AA) from S-VWN, S-LYP, B-VWN, and B-LYP
calculations on the transition metal fluoride series. The calculations
were carried out with
the 6-31G+* basis on the fluorine atoms and the $[6s5p3d]$
contraction described in the text. The experimental results are
summarized in Ref.~\protect\citenum{DH}.}
\label{tb:results3}
\end{table}

\begin{table}[ht]
\begin{center}
\begin{tabular}{llr|r|r||r|r|r||r}  \hline\hline
&&\multicolumn{3}{c}{$[6s5p3d]$}&
  \multicolumn{3}{c}{$[6s5p3d1f]$}&
  \multicolumn{1}{c}{Expt}\\
\multicolumn{2}{l}{Molecule}
                        &HF   &S-VWN&B-LYP&HF   &S-VWN&B-LYP&     \\  \hline
$\mbox{ScF}_3$ &        &133.8&175.6&154.4&135.7&177.3&155.6&147.4(147.7)\\
$\mbox{TiF}_4$ &        & 91.2&171.5&147.8& 94.1&173.6&149.5&139.9(140.4)\\
$\mbox{VF}_5$  &        & 49.4&148.8&123.6& 53.3&151.2&125.5&112.1(112.7)\\
$\mbox{CrF}_6$ &${O_h}$   &     &     & 92.6&     &     &94.5&     \\
               &${D_{3h}}$&     &     & 89.9&     &     &91.7&     \\
\hline\hline
\end{tabular}
\end{center}
\caption{Average bond energies $\overline {E}$(kcal/mol)
from HF, S-VWN, and B-LYP calculations on the transition metal
fluoride series.  The calculations were carried out with the 6-31G+*
basis on the fluorine atoms and either the $[6s5p3d]$ or $[6s5p3d1f]$
contraction described in the text.  The experimental values are
obtained from heats of formation at
298${}^\circ$K\protect\cite{barin}.  Estimates of the experimental
values ``corrected'' for zero-point energy and enthalpy differences
between 0 and 298${}^\circ$K are given parenthetically (see text);
these may be compared directly to the theoretical results.}
\label{tb:results2}
\end{table}

\begin{table}[ht]
\begin{center}
\begin{tabular}{llrr|rr}\hline\hline
              &
&\multicolumn{2}{c}{Becke3LYP}&\multicolumn{2}{c}{Experiment}\\
\multicolumn{2}{l}{Molecule}&$r_{eq}$&$\overline{E}$&$r_{eq}$&$\overline{E}$\\\hline
$\mbox{ScF}_3$&       &1.852&149.7&1.91 &147.4(147.7)\\
$\mbox{TiF}_4$&       &1.756&141.8&1.745&139.8(140.4)\\
$\mbox{VF}_5$&$r_{ax}$&1.750&114.4&1.734&112.1(112.7)\\
             &$r_{eq}$&1.717&     &1.703&     \\
$\mbox{CrF}_6$&$O_h$&1.732&78.8&&\\
$\mbox{CrF}_6$&$D_{3h}$&1.743&76.5&&\\
\end{tabular}
\end{center}
\caption{Equilibrium bond lengths and average bond energies calculated
with Becke3LYP method compared to experimental values.  The
$[6s5p3d1f]$/6-31G+* basis set was used.  Bond energies were
calculated using UKS energies for atoms.  The experimental values are
obtained from Ref.~\protect\citenum{DH} and
Ref.~\protect\citenum{barin}; see caption for Table~\protect\ref{tb:results2}
and text.}
\label{tb:b3lyp}
\end{table}

\begin{table}[ht]
\begin{center}
\begin{tabular}{l||l||l||l}  \hline\hline
         & B-LYP                     &  Becke3LYP      & Experiment\\
Mode     &$\omega_e (\mbox{cm}^{-1})$&$\omega_e (\mbox{cm}^{-1})$  \\  \hline
$t_{2u}$ &157.8&145.2&\\
$t_{1u}$ &329.0&335.7(27.06)& 332(0.08)\\
$t_{2g}$ &344.2&363.3&\\
$e_{g} $ &554.7&591.8&\\
$a_{1g}$ &648.8&708.4&\\
$t_{1u}$ &717.0&767.7(333.67)& 760(1.0)\\ \hline \hline
\end{tabular}
\end{center}
\caption{Vibrational frequencies in $\mbox{CrF}_6$.
The calculations were carried out with the 6-31G+* basis on the
fluorine atoms and the $[6s5p3d1f]$ metal contraction described in the
text. They were obtained by finite differencing analytic gradients at
the $O_h$ equilibrium geometry.  Numbers in parentheses are the IR
intensities in KM/mol.  The experimental frequencies and relative
intensities are from Ref.~\protect\citenum{JW}.}
\label{tb:freqs}
\end{table}

%\begin{table}[ht]
%\begin{center}
%\begin{tabular}{lr}\hline\hline
%Molecule&Binding energy\\\hline
%$\mbox{ScF}_3$&442.2\\
%$\mbox{TiF}_4$&559.4\\
%$\mbox{VF}_5$&560.4\\\hline\hline
%\end{tabular}
%\end{center}
%\caption{Experimental values for binding energies, obtained from
%Ref.~\protect\citenum{barin}.}
%\label{tb:expt}
%\end{table}
\clearpage

\section{Figure Captions}

\end{document}